\documentclass[11pt]{article}
\usepackage{graphicx}

\begin{document}

\title{Maximal Information Transfer and Behavior Diversity in Random Threshold Networks}

\author{M. Andrecut, D. Foster, H. Carteret and S. A. Kauffman}

\date{ }

\maketitle

{\par\centering Institute for Biocomplexity and Informatics \par}
{\par\centering University of Calgary \par}
{\par\centering 2500 University Drive NW, Calgary \par}
{\par\centering Alberta, T2N 1N4, Canada \par}

\noindent

\begin{abstract}

Random Threshold Networks (RTNs) are an idealized model of di-
luted, non-symmetric spin glasses, neural networks or gene regulatory
networks. RTNs also serve as an interesting general example of any
coordinated causal system. Here we study the conditions for maximal
information transfer and behavior diversity in RTNs. These conditions
are likely to play a major role in physical and biological systems, per-
haps serving as important selective traits in biological systems. We
show that the pairwise mutual information is maximized in dynami-
cally critical networks. Also, we show that the correlated behavior di-
versity is maximized for slightly chaotic networks, close to the critical
region. Importantly, critical networks maximize coordinated, diverse
dynamical behavior across the network and across time: the informa-
tion transmission between source and receiver nodes and the diversity
of dynamical behaviors, when measured with a time delay between
the source and receiver, are maximized for critical networks.
\bigskip \bigskip

\end{abstract}
\pagebreak

\section{Introduction}

The random Boolean networks (RBNs) model was initially introduced as an 
idealized model of genetic regulatory networks [1]. Since then, the RBN model 
has attracted much interest in a wide variety of fields, ranging from cell 
differentiation and evolution to social and physical spin systems. The dynamics 
of RBNs can be classified as ordered, chaotic, or critical, as a function of 
the average in-degree $k$ and the bias $p$ in the choice of Boolean functions [2]. 
The central issue of the research on the RBN model is the characterization of 
the critical transition between ordered and chaotic phases [3]. These two regimes 
produce very different emergent dynamical behaviors. Networks operating in 
the ordered regime are intrinsically robust, but exhibit simple dynamics. 
This robustness is reflected in the dynamical stability of the network both 
under structural perturbations and transient perturbations. In contrast, 
networks in the chaotic regime are extremely sensitive to small perturbations, 
which rapidly propagate throughout the entire system. The phase transition 
between the ordered and chaotic regimes represents a tradeoff between stability 
and access to a wide range of dynamic behavior to respond to a variable
environment. 

Recently it has been shown that the pairwise mutual information exhibits 
a jump discontinuity at the critical values of $k$ and $p$ [4]. Here, 
we extend these results to a second class of discrete dynamical networks 
called Random Threshold Networks (RTN), which were first studied as
diluted, non-symmetric spin glasses, neural networks and gene regulatory 
networks [5-9].  More specifically, we study the conditions for maximal 
information transfer and behavior diversity in RTNs. We show that the 
pairwise mutual information is maximized in critical networks. Also, 
we show that the correlated behavior diversity is maximized for slightly 
chaotic networks, close to the critical region when measured with no 
delay between source and receiver nodes. In contrast, when the delay 
between the measurement of source and receiver nodes is increased, 
correlated behavior complexity and diversity is maximized for critical 
networks. These results support the hypotheses that critical networks 
provide an optimal information transfer between the elements of the 
network, and optimal coordination of diverse behavior when there is 
a time delay between source and receiver nodes, while slightly chaotic 
networks have an optimal capacity for coordinating most diverse dynamical 
behavior in the absence of such a delay.

\section{RTN model}
The RTN model consists of $N$ randomly interconnected binary variables
(spins) with states $\sigma _{i}\in \{\pm 1\}$, $i=0,...,N-1$ [9]. Each
variable has associated a function: 
\begin{equation}
\varphi _{i}(t)=\sum_{j=0}^{N-1}\omega _{ij}\sigma _{j}(t)+\theta ,
\end{equation}
where the interaction weights take discrete values $\omega _{ij}\in \{\pm 1\}
$ with equal probability, and the discrete threshold $\theta $ is fixed.
Without loosing generality, in the following discussion the threshold
parameter is set to $\theta =0$. If node $i$ does not receive signals from
node $j$, one has $\omega _{ij}=0$. The average number $k$ of non-zero
interaction weights represents the average connectivity (in-degree) of the
nodes in the network. The variable $\sigma _{i}$ will change its state at
each time step according to the rule: 
\begin{equation}
\sigma _{i}(t+1)=sign(\varphi _{i}(t)).
\end{equation}
where $sign(x)=1$ if $x\geq 0$ and $sign(x)=-1$ if $x<0$.

\section{Information transfer and behavior diversity}
The average pairwise correlation has been used to characterize the typical 
dynamics of pairs of nodes [9]. The average correlation between two nodes $i$ and $j$ is defined as: 
\begin{equation}
C_{ij}=\left| \frac{1}{T}\sum_{t=0}^{T-1}\sigma _{i}(t)\sigma _{j}(t)\right|
,
\end{equation}
where $T$ is the length of the time series over which the correlation is measured. 
If the dynamical activity of two nodes $i$ and $j$ is (anti-)correlated, i.e. if $\sigma
_{i}(t)$ and $\sigma _{j}(t)$ always have either the same or the opposite
sign, one has $C_{ij}=1$. If the relationship between the signs of $\sigma
_{i}(t)$ and $\sigma _{j}(t)$ occasionally changes then $0\leq C_{ij}<1$. It
has been shown that the quantity: 
\begin{equation}
\left\langle C\right\rangle =N^{-2}\left\langle C_{ij}\right\rangle ,
\end{equation}
(averaged over the RTN ensemble) exhibits a second order phase transition at
a critical average connectivity $k_{c}=2$ [9]. For $k<k_{c}$, the nodes are
typically frozen, with $\left\langle C_{ij}\right\rangle \simeq 1$. For $%
k\geq k_{c}$, in the limit of large $N$, $\left\langle C_{ij}\right\rangle $
undergoes a transition at $k_{c}$, vanishing for larger $k>>k_{c}$ [9].

Now, let us define the average activity $A_{i}$ of a node $i$ as following: 
\begin{equation}
A_{i}=1-\left| \frac{1}{T}\sum_{t=0}^{T-1}\sigma _{i}(t)\right| ,
\end{equation}
Frozen nodes, which do not change their states have an activity $A_{i}=0$,
while the nodes who occasionally change their state have an activity $%
0<A_{i}\leq 1$. The average activity 
\begin{equation}
\left\langle A\right\rangle =N^{-1}\left\langle A_{i}\right\rangle
\end{equation}
is largest $\left\langle A\right\rangle \sim 1$ in the chaotic phase $%
k>k_{c} $, and by decreasing $k$ undergoes a second order phase transition
at $k_{c}$, vanishing for $k=0$.

We can define the entropy [10] of the node $i$, $i=0,1,...,N-1$, as: 
\begin{equation}
H_{i}=-\sum_{\alpha \in \{\pm 1\}}p_{\alpha }(\sigma _{i}(t))\log
_{2}p_{\alpha }(\sigma _{i}(t)),
\end{equation}
where $p_{\alpha }(\sigma _{i}(t))$ is the probability of the symbol $\alpha
\in \{\pm 1\}$ in the time series $\{\sigma _{i}(t)\}$, $i=0,1,...,N-1$. The
average entropy of the RTN ensemble is: 
\begin{equation}
\left\langle H\right\rangle =N^{-1}\left\langle \sum_{i}H_{i}\right\rangle .
\end{equation}
Both, the activity $A$ and the entropy $H$ measure the diversity of the
dynamics of the nodes in the RTN ensemble as a function of their
connectivity $k$.

The mutual information [11] between the nodes $i$ and $j$ as a function of
the time lag $\tau =0,1,2,...$ is defined as following: 
\begin{equation}
I_{ij}(\tau )=\sum_{\alpha \in \{\pm 1\}}\sum_{\beta \in \{\pm 1\}}p_{\alpha
\beta }(\sigma _{i}(t),\sigma _{j}(t+\tau ))\log _{2}\left( \frac{p_{\alpha
\beta }(\sigma _{i}(t),\sigma _{j}(t+\tau ))}{p_{\alpha }(\sigma
_{i}(t))p_{\beta }(\sigma _{j}(t))}\right) .
\end{equation}
Here, $p_{\alpha }(\sigma _{i}(t))$ and $p_{\beta }(\sigma _{i}(t+\tau ))$
are the probabilities of the symbols $\alpha ,\beta \in \{\pm 1\}$ in $%
\{\sigma _{i}(t)\}$ and respectively $\{\sigma _{j}(t)\}$, and $p_{\alpha
\beta }(\sigma _{i}(t),\sigma _{j}(t+\tau ))$ is the probability of the pair 
$(\alpha ,\beta )$ in $\{(\sigma _{i}(t),\sigma _{j}(t+\tau ))\}$. $%
I_{ij}(\tau )$ measures the extent to which information about node $i$ at
time $t$ influences node $j$ at time $t+\tau $. The propagation of
information may be indirect, i.e. both nodes are influenced by a common node
through previous time steps. The above probabilities can be easily estimated
as following: 
\begin{equation}
p_{\alpha \beta }(\sigma _{i}(t),\sigma _{j}(t+\tau ))=\frac{%
\sum_{t=0}^{T-\tau }\delta (\sigma _{i}(t);\alpha )\delta (\sigma
_{j}(t+\tau );\beta )}{\sum_{\alpha \in \{\pm 1\}}\sum_{\beta \in \{\pm
1\}}\sum_{t=0}^{T-\tau }\delta (\sigma _{i}(t);\alpha )\delta (\sigma
_{j}(t+\tau );\beta )},
\end{equation}
\begin{equation}
p_{\alpha }(\sigma _{i}(t))=\sum_{\beta \in \{\pm 1\}}p_{\alpha \beta
}(\sigma _{i}(t),\sigma _{j}(t+\tau )),
\end{equation}
\begin{equation}
p_{\beta }(\sigma _{i}(t))=\sum_{\alpha \in \{\pm 1\}}p_{\alpha \beta
}(\sigma _{i}(t),\sigma _{j}(t+\tau )),
\end{equation}
where 
\begin{equation}
\delta (x;y)=\left\{ 
\begin{array}{lll}
1 & if & x=y \\ 
0 & if & x\neq y
\end{array}
\right. ,
\end{equation}
is the Dirac delta function, and $T$ is the length of the considered time
series. In order to characterize the information propagation through the
entire network, we define the average pairwise mutual information for the
RTN ensemble as following: 
\begin{equation}
\left\langle I(\tau )\right\rangle =N^{-2}\left\langle \sum_{i,j}I_{ij}(\tau
)\right\rangle .
\end{equation}
Because the number of pairs $(i,j)$ that contribute significantly to the sum
is expected to be at most of order $N$, it is convenient to work with the
quantity 
\begin{equation}
I_{N}(\tau )=N\left\langle I(\tau )\right\rangle ,
\end{equation}
which approaches a nonzero constant in the large-$N$ limit [4].

One can define the pairwise correlated behavior in several ways:

\begin{enumerate}
\item  as the product between the activity of the nodes and their
correlation: 
\begin{equation}
D_{ij}=\frac{1}{2}(A_{i}+A_{j})C_{ij},
\end{equation}
with the ensemble average given by: 
\begin{equation}
\left\langle D\right\rangle =N^{-2}\left\langle D_{ij}\right\rangle .
\end{equation}

\item  as a product between the entropy of the nodes and their correlation: 
\begin{equation}
F_{ij}=\frac{1}{2}(H_{i}+H_{j})C_{ij},
\end{equation}
with the ensemble average given by: 
\begin{equation}
\left\langle F\right\rangle =N^{-2}\left\langle F_{ij}\right\rangle .
\end{equation}

\item  as the product between the entropy of the nodes and their mutual
information: 
\begin{equation}
Q_{ij}(\tau )=\frac{1}{2}(H_{i}+H_{j})I_{ij}(\tau ),
\end{equation}
with the ensemble average given by: 
\begin{equation}
Q_{N}(\tau )=\frac{1}{2}N^{-1}\left\langle (H_{i}+H_{j})I_{ij}(\tau
)\right\rangle .
\end{equation}
\end{enumerate}

Obviously, these quantities measure the correlated behavior diversity.
For example, $D_{ij}$ and $F_{ij}$ are high only if both the activity
(entropy) of the nodes and their correlation are simultaneously high. 
This means both that the nodes are changing, and that they are changing 
in a correlated fashion. Also, assuming that mutual information is a 
correlation measure, we may say that $Q_{ij}(\tau )$ measures the diversity 
of correlated dynamical behavior between the present $t$ and the future 
at $t+\tau $. Thus, by locating the ensemble of networks with a connectivity 
$k$ which maximizes these measures, we have found the dynamical region which 
exhibits the most correlated behavior diversity in the present (for $D$, $F$), 
and between the present and the future (for $Q(\tau)$).

\section{Numerical results}

Since the above defined quantities are not yet analytically accessible, numerical
simulations are necessary for their estimation. The simulation procedure is
similar to the one described in [3, 4]. For networks of size $N=10^{3}$, we simulated the dynamics for $10^3$ steps to eliminate the transient dynamics, 
and collected a time series of length 
$T=10^{4}$. Also, we averaged over $10^{3}$ distinct, randomly generated networks with $100$ runs from
different randomly chosen initial states for each network. We should note that the 
ensemble averages are computed using a Monte Carlo procedure which includes data from all the attractors generated in
the calculation. For example, in order to calculate $\left\langle C\right\rangle $ we sample  $C_{i,j}$ 
randomly over $S=10^{3}$ networks and $R=100$ runs from different initial states, and we discard the transient dynamics for each run. 
The sampling is done by computing $C_{i,j}$ for $M=100N$ randomly generated pairs $m\equiv random(i,j)$, in each run. 
All the obtained samples $C_{m}$ are then used to calculate an approximation 
of the ensemble average as: 
\begin{equation}
\left\langle C\right\rangle \simeq  \frac{1}{S R M} \sum_{s=1}^{S} \sum_{r=1}^{R} \sum_{m=1}^{M} C_{m}.
\end{equation}

In Figure 1 we give the average activity $\left\langle A\right\rangle $, the
average correlation $\left\langle C\right\rangle $, the entropy $\left\langle H\right\rangle $ 
and their numerical derivatives with respect
to $k$: $d\left\langle A\right\rangle /dk$, $\left| d\left\langle C\right\rangle /dk \right|$, $d\left\langle H\right\rangle /dk$. One can see that all three quantities
exhibit a phase transition around $k_{c}^{\prime }\simeq 2.2$. The deviation from the large-network-size limit
value of $k_{c}=2$, when $N\rightarrow \infty $, is due to the finite size of the simulated networks: 
$\Delta k_{c}=k_{c}^{\prime }-k_{c}\simeq 0.2$. 

The average correlated behavior defined by  $\left\langle D\right\rangle $ and $\left\langle F\right\rangle $ is shown in
Figure 2. Both $\left\langle D\right\rangle $ and $\left\langle F\right\rangle 
$ reach their maximum value in the slightly chaotic regime, around $k_{DF}^{\prime
}\simeq 2.6$ ($k_{DF}\simeq 2.4$, corrected for the finite size effect, as noted above). 

In Figure 3 we show the average mutual information $I_{N}(\tau )$ as a function
of the time lag $\tau =0,1,...,40$. The mutual information also has a maximum
value around the critical value $k_{c}^{\prime }\simeq 2.2$ ($k_{c}\simeq 2.0
$, corrected for the finite size effect). It is interesting to note that $%
I_{N}(\tau )$ decreases with $\tau $, having a maximum value for $\tau =0$. 
This shows that information gained about the state of one node given the state of 
another node decays over time, as one might expect. In addition, the numerical 
simulation suggests that $I_{N}(\tau )$ is localizing
around the critical value $k_{c}$ when the time lag $\tau $ increases,
converging to a delta function for large $\tau $: $I_{N}(\tau )\rightarrow
\delta (k,k_{c})$. 

The average correlated behavior $Q_{N}(\tau )$ is given in Figure 4. 
The maximum of $Q_{N}(\tau )$ shifts to the left when $\tau $ increases, from $k_{c}^{\prime
}\simeq 2.4$ when $\tau =0$, to $k_{c}\simeq 2.2$ when $\tau =40$, and it
seems to converge to $k_{c}$ for large $\tau $. 

\section{Discussion and conclusion}

Shannon information measures the information transmission down a noisy channel with a 
decoder of indefinite computational power and seeks to maximize information transmission [10]. 
Cells, other biological and other physical systems, do not have decoders of arbitrary power. 
More, it seems plausible that in cells, neural systems, and other tissues, natural 
selection will have acted to maximize both information transfer across the network, 
and the diversity of complex behaviors that can be coordinated within a causal network. 
We have shown that the pairwise mutual information is maximized in critical RTNs. 
Also, we have shown that the diversity of complex correlated behavior is maximized 
for slightly chaotic RTNs, close to the critical region using two measures of correlated 
diversity, $D$ and $F$, with no temporal delay between signaling and receiving nodes.  
Importantly, in the presence of a delay, $\tau$, between signaling and receiving nodes, 
maximum diversity of complex coordinated behaviors clearly shifts towards critical 
networks as the delay increases. Ordered networks have convergent trajectories, and hence "forget" their past; chaotic networks show sensitivity 
to initial conditions, and thus they, too, forget their past and are 
unable to act reliably. Critical networks, with trajectories that, on average, neither 
diverge or converge, seem best able to bind past to future.  In short, our results show that in 
the presence of a delay, hence time binding, critical networks maximize information transfer 
between source and receiver nodes, i.e. they maximize pairwise mutual information, and simulataneously 
maximize the diversity and complexity of behaviors that can be correlated by that information transfer.

Given the potential biological implications, it is of interest that recent data suggest 
that genetic regulatory networks in eukaryotic cells are dynamically critical [12-14].
Also, recent experiments conducted on rat brain slices show that these neural tissues are critical [15].  
RTN are simple Boolean models of threshold neural networks. Further work with random 
Boolean networks, RBN, will attempt to extend these results to this class of disordered 
causal systems, and will extend these results to communication between networks.  
We note that recent results have shown that critical RBN maximize power efficiency [3, 16].  
Maximum energy efficiency occurs if work cycles are performed infinitely slowly. 
Cells must do work cycles to reproduce.  Infinitely slow cell reproduction would fail 
in the Darwinian race. Maximum power efficiency occurs at a finite, defined, displacement 
from equilibrium.
Our hope is that subsequent work will establish that cells and tissues, 
as evolved, far from equilibrium evolved systems, simultaneously maximize information transfer, 
the complexity and diversity of dynamical behaviors that can be coordinated, and the power 
efficiency with which these complex diverse behaviors are carried out. Such results may help formulate a 
far from equilibrium theory for living systems.

Using random threshold boolean nets as simple models of complex causal systems we have studied 
information transfer between source and receiver nodes.  We have shown that critical 
RTN maximize both information transfer, and the complexity and diversity of dynamical behaviors 
that can be coordinated across the network and time.

\clearpage
\begin{figure}
\centering
\includegraphics[width=9cm]{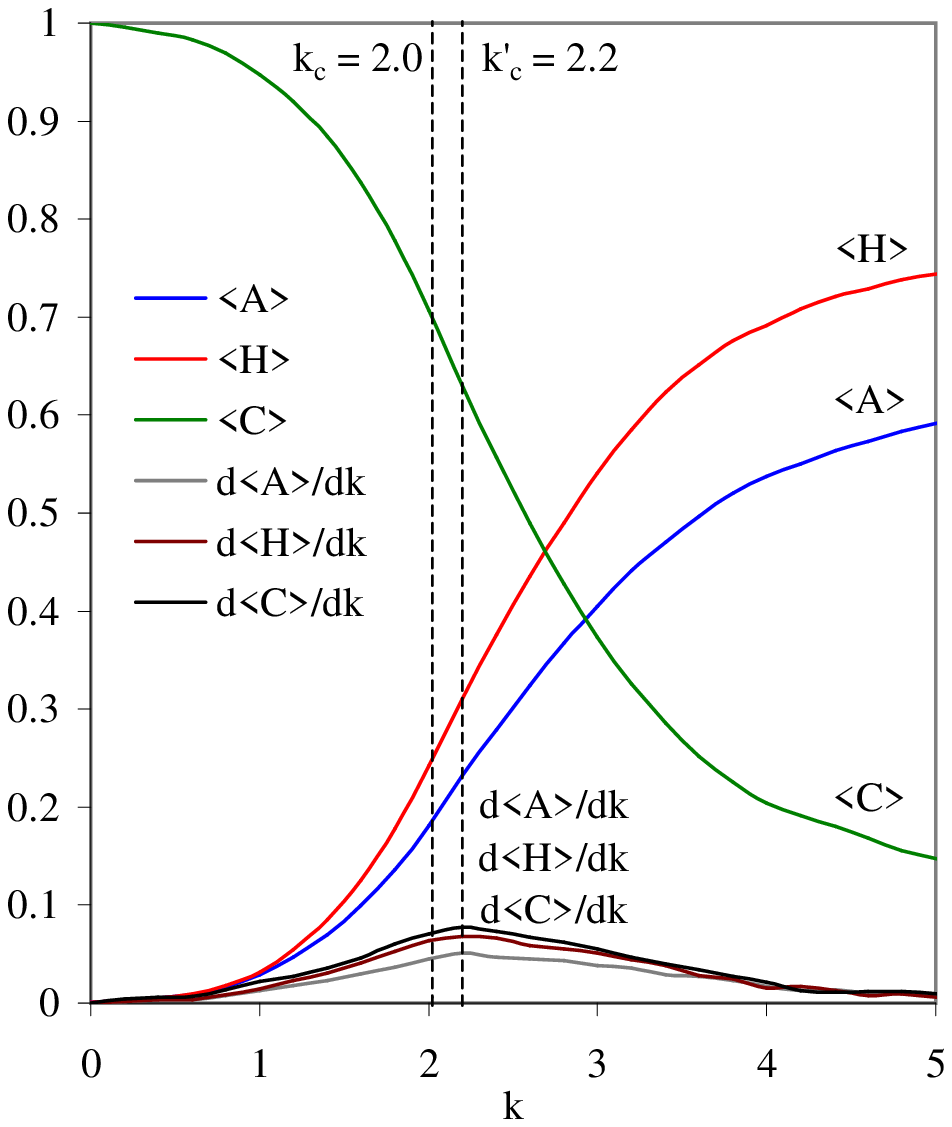}
\caption{\label{Fig1} The average activity $\left\langle A\right\rangle $, the
average correlation $\left\langle C\right\rangle $, the entropy $\left\langle H\right\rangle $ 
and their numerical derivatives with respect
to $k$: $d\left\langle A\right\rangle /dk$, $d\left\langle C\right\rangle /dk
$, $d\left\langle H\right\rangle /dk$. }
\end{figure}

\clearpage
\begin{figure}
\centering
\includegraphics[width=9cm]{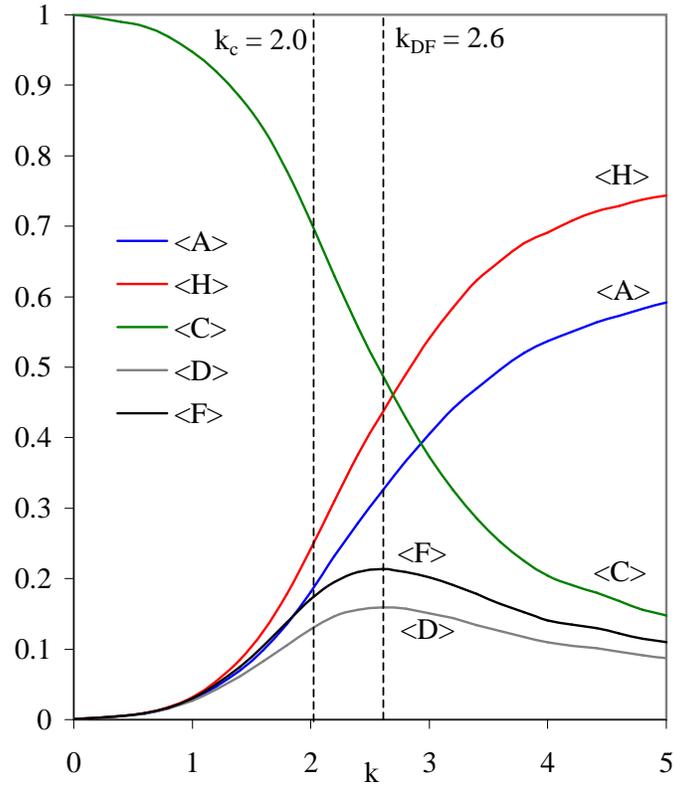}
\caption{\label{Fig2} The average activity $\left\langle A\right\rangle $, the
average correlation $\left\langle C\right\rangle $, the entropy $\left\langle H\right\rangle $ 
and the average correlated behavior defined by $\left\langle D\right\rangle $ and $\left\langle F\right\rangle $.}
\end{figure}

\clearpage
\begin{figure}
\centering
\includegraphics[width=9cm]{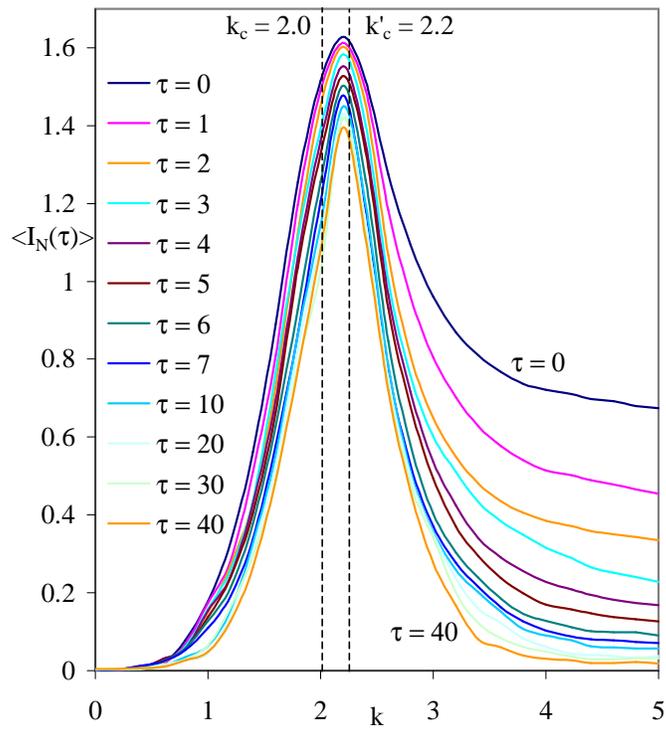}
\caption{\label{Fig3} The average mutual information $I_{N}(\tau )$ as a function
of the time lag $\tau =0,1,...,7$.}
\end{figure}

\clearpage
\begin{figure}
\centering
\includegraphics[width=9cm]{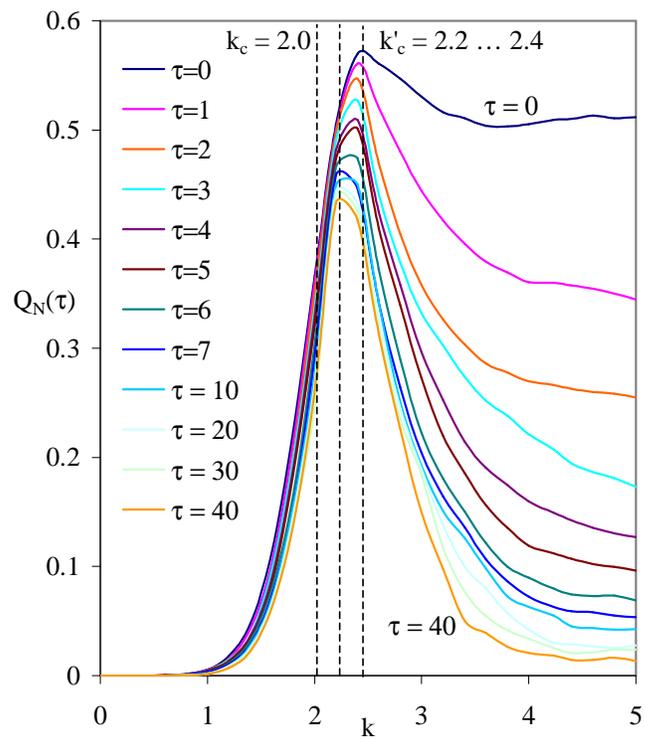}
\caption{\label{Fig4} The average correlated behavior $Q_{N}(\tau )$ as a function
of the time lag $\tau =0,1,...,7$.}
\end{figure}

\end{document}